\documentclass[12pt,preprint]{aastex}

\usepackage{natbib}
\usepackage[usenames]{color}
\title{Numerical examination of plasmoid-induced reconnection model for solar flares: the relation between plasmoid velocity and reconnection rate}
\author{Keisuke Nishida\altaffilmark{1}, Masaki Shimizu\altaffilmark{2}, Daikou Shiota\altaffilmark{3}, Hiroyuki Takasaki\altaffilmark{4}, Tetsuya Magara\altaffilmark{3}, and Kazunari Shibata\altaffilmark{1}}

\altaffiltext{1}{Kwasan and Hida Observatories, Kyoto University, Yamashina, 
	Kyoto 607-8471, Japan; nishida@kwasan.kyoto-u.ac.jp}
\altaffiltext{2}{Department of Mechanical Engineering and Science, Kyoto University,
Yoshida-Honmachi, Sakyo, Kyoto 606-8501, Japan.}
\altaffiltext{3}{National Astronomical Observatory, Mitaka, Tokyo 181-8588, Japan.}
\altaffiltext{4}{Accenture Japan Ltd, Akasaka Inter City, 1-11-44 Akasaka Minato-ku, Tokyo 107-8672, Japan.}

\shortauthors{Nishida et al.}

\begin{document}

\begin{abstract}
The plasmoid-induced-reconnection model explaining solar flares based on bursty reconnection produced by an ejecting plasmoid suggests a possible relation between the ejection velocity of a plasmoid and the rate of magnetic reconnection. In this study, we focus on the quantitative description of this relation. We performed magnetohydrodynamic (MHD) simulations of solar flares by changing the values of resistivity and the plasmoid velocity.
The plasmoid velocity has been changed by applying an additional force to the plasmoid to see how the plasmoid velocity affects the reconnection rate.
An important result is that the reconnection rate has a positive correlation with the plasmoid velocity, which is consistent with the plasmoid-induced-reconnection model for solar flares.
We also discuss an observational result supporting this positive correlation.
\end{abstract}


\section{Introduction}\label{introduction}
Magnetic reconnection is a process in which the magnetic energy is converted into the kinetic and thermal energy \citep{1958IAUS....6..123S, 1963ApJS....8..177P, 1964psf..conf..425P}, and it has been widely believed to play a fundamental role in causing solar flares. 
A model for flares based on magnetic reconnection has been developed since 1960s by 
\citet{1964psf..conf..451C}, \citet{1966Natur.211..697S}, \citet{1974SoPh...34..323H},
and \citet{1976SoPh...50...85K}, so this model has been called the CSHKP model.
The observations supporting this model has been reported,
such as cusps \citep{1992PASJ...44L..63T}, arcades \citep{1992PASJ...44L.211T, 1992PASJ...44L.205M, 2002GeoRL..29u..10I}, loop top hard X-ray (HXR) sources \citep{1994Natur.371..495M,2003ApJ...596L.251S}, X-ray jets \citep{1992PASJ...44L.173S, 1996PASJ...48..123S}, and so on.
Furthermore Yohkoh \citep{1991SoPh..136....1O} discovered the common property of flares with different appearances, such as Long Duration Events (LDE flares) and impulsive flares (see \citealt{1999Ap&SS.264..129S} and \citealt{2002SSRv..101....1A} for review).
%%LDE flares and impulsive flares are distinguished with the duration of X-ray 
%%brightening; The duration of LDE flares is typically longer than one hour and the 
%%duration of impulsive flares is shorter than one hour.
%%The discoveries lead us to have a view that unifies the two types of flares.

So far the dynamic process caused by magnetic reconnection in flares has been widely studied by MHD simulations \citep[e.g.][]{1983SoPh...84..169F, 1990JGR....9511919F, 1991SoPh..135..361F, 1996ApJ...466.1054M, 1996PhPl....3.4172U, 2000JGR...105.2375L, 2001ApJ...549.1160Y}.
Most of these works have been focused on MHD processes producing apparent features of flares such as cusp shaped loops, ejecting plasmoid (blob of plasma), inflows and loop-top HXR sources.
However, these works have not clearly explained the fundamental physical process, that is what determines the energy release rate in flares. This question is related to the rate of magnetic reconnection, so to identify the condition under which fast magnetic reconnection \citep{1977JPlPh..17..337U} operates is important.

%%It is known that localized resistivity triggers Petscheck-type fast reconnection, \citep{1977JPlPh..17..337U}.
%%The so-called anomalous resistivity satisfies this condition, but how this bursty reconnection occurs was unclear.

Observationally, solar flares are often associated with plasmoid ejections.
\citet{1995ApJ...451L..83S} found that 8 impulsive flares on the limb (Masuda-type; \citealt{1995PASJ...47..677M}) were associated with plasmoid ejections. \citet{1997PASJ...49..249O, 1998ApJ...499..934O} carefully analyzed plasmoid ejections in impulsive flares and found that plasmoids undergo strong acceleration during the impulsive phase of these flares.
\citet{2004ApJ...616..578S} found the tiny two-ribbon flare driven by emerging flux accompanying the miniature filament eruption (=plasmoid).
This feature was also found in other observations \citep{2001ApJ...559..452Z,2004ApJ...613..592T,2005ApJ...630.1148S} and numerical simulations \citep{1997ApJ...487..437M}.
It is also found that there is a positive correlation between the plasmoid velocity and the reconnection rate
\citep{1995ApJ...451L..83S, 2005ApJ...634L.121Q, shimizu2008}.
There are other literatures discussing this topic \citep{1999ApJ...525L..57N,2000JGR...10523153F, Klimchuk2001, 2002A&ARv..10..313P,2003NewAR..47...53L,2005ApJ...635.1291K}.
\citet{2000JGR...105.2375L} derived an analytic relation between the acceleration of coronal mass ejections (CMEs) and reconnection rate.

The observations above show an important suggestion that plasmoid ejection plays a key role in causing fast magnetic reconnection. Based on these observational results, \citet{1996AdSpR..17....9S, 1997cswn.conf..103S} extended the classical CSHKP model and proposed the plasmoid-induced reconnection model. In this model, a plasmoid (or flux rope in a 3D situation) is created in the anti-parallel magnetic field by the magnetic reconnection (Figure \ref{figure:pir}a, b, c). Then the plasmoid situates in a current sheet inhibits inflows into the sheet, so reconnection is inefficient and magnetic energy is stored (Figure \ref{figure:pir}d). 
%%We can consider the two types of plasmoids to be same, the plasmoid newly created in the current sheet by the reconnection (Figure \ref{figure:pir}a) and the plasmoid which is preexists before the start of the reconnection (Figure \ref{figure:pir}e).
Then the plasmoid starts to move at the velocity $v_{plasmoid}$, inflows toward the X-point ($v_{inflow}$) are induced following mass conservation and reconnection starts (Figure \ref{figure:pir}e).
If we assume the incompressibility, the mass flux into the reconnection region is given by $\sim v_{inflow} L_{inflow}$ and the mass flux ejected by the plasmoid  is $\sim v_{plasmoid} W_{plasmoid}$, and these are balanced, where $W_{plasmoid}$ is the typical width of the plasmoid, and $L_{inflow}$ $(\ge W_{plasmoid})$ is the typical length of the inflow region.
Consequently the induced inflow speed can be estimated as follows:
\begin{equation}
v_{inflow} \sim v_{plasmoid} W_{plasmoid} / L_{inflow}.
\end{equation}
Since the reconnection rate is determined by the speed of the inflows, fast reconnection becomes possible when plasmoid ejection is fast.
Moreover the jet from a reconnection point accelerate the plasmoid
\footnote{Note that in addition to the acceleration by the reconnection jet, the magnetic pressure gradient force can also accelerate the plasmoid if there is a global magnetic pressure gradient around the plasmoid. See more detailed discussion in Appendix.},
 so the fast reconnection further drives fast plasmoid ejection.
This suggests a positive correlation between the plasmoid velocity and the reconnection rate. The merit of this model is to provide us with a unified view for understanding various types of flares with different spatial sizes and timescales, and it can naturally cause fast reconnection \citep{1986JGR....91.5579P} in a current sheet where many plasmoids with different sizes are created by tearing instability \citep{2000A&A...360..715K,2001EP&S...53..473S,2004A&A...417..325K}.

\begin{figure}
\plotone{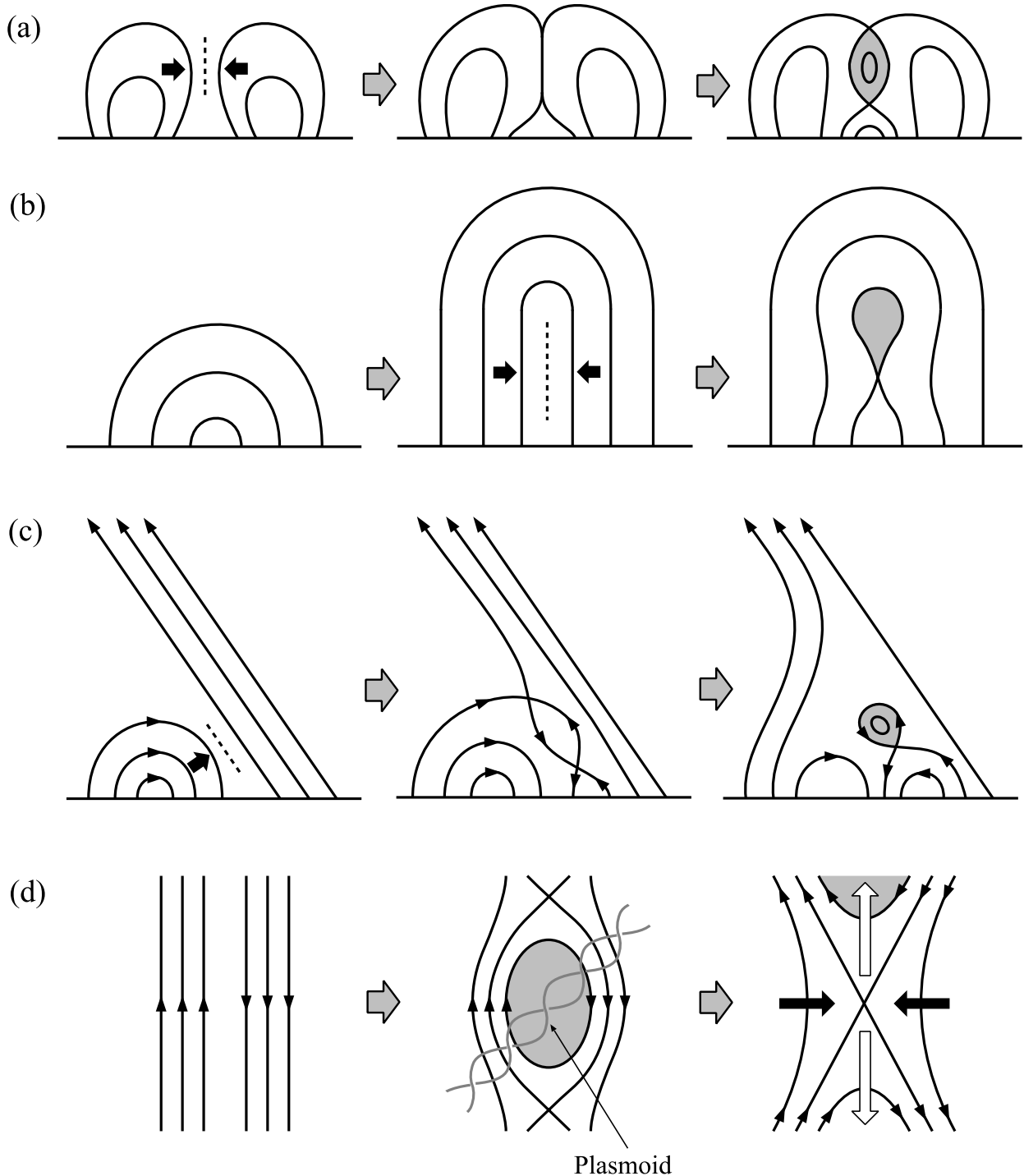}

\caption{Schematic picture of the plasmoid-induced reconnection model. Solid lines indicate magnetic field lines. Panel (a), (b) and (c) shows the process creating the plasmoid in the anti-parallel magnetic field by the magnetic reconnection.
Panel (d) shows how the plasmoid in the current sheet inhibits the magnetic reconnection  in the reconnection region shown by a light gray area in panel (e).
Thick black arrows indicate inflows to a reconnection point, and thick white arrows indicate reconnection jets. Panel (e) explains how the plasmoid ejection can induce strong inflows into the reconnection region.}
\label{figure:pir}
\end{figure}

\setcounter{figure}{0}
\begin{figure}
\plotone{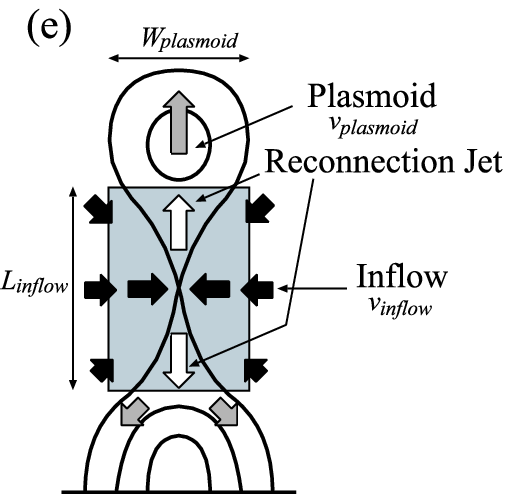}
\caption{cont.}
\end{figure}

So far there have been no MHD simulations performed to examine the plasmoid-induced-reconnection model,
which should reproduce the correlation between the plasmoid velocity and the reconnection rate.
In this study, we performed a series of these MHD simulations by changing the parameters related to resistivity and plasmoid velocity to investigate the relation between these two quantities.

The model and the numerical method are described in section \ref{section:model}, and the numerical results are presented in section \ref{section:results}. The discussion is given in section \ref{section:discussion} and the conclusions are given in section \ref{section:conclusion}.

\section{Numerical Method and Model}\label{section:model}

\subsection{Numerical Method}

We performed simulations with a multistep implicit scheme \citep{1989JCoPh..84..441H} in order to solve the 2.5-dimensional time-dependent resistive MHD equations in the Cartesian coordinates ($x$, $y$, $z$) where $y$ is directed upward:

\begin{equation}
\frac{\partial \rho}{\partial t} + \nabla \cdot (\rho \mathbf{v}) = 0,
\end{equation}

% Equation of Motion
\begin{equation}
\frac{\partial\mathbf{v}}{\partial t}+(\mathbf{v}\cdot\nabla)\mathbf{v} + \frac{1}{\rho}\nabla p - \frac{1}{\rho}\mathbf{j}\times\mathbf{B}-\mathbf{F}=0, \label{fqn:motion}
\end{equation}

\begin{equation}
\frac{\partial\mathbf{B}}{\partial t} + \nabla\times(\mathbf{v}\times\mathbf{B}) - \eta\nabla\times \mathbf{j} = 0,
\end{equation}

\begin{equation}
\frac{\partial T}{\partial t} + \mathbf{v} \cdot \nabla T + (\gamma - 1)T \nabla \cdot \mathbf{v} - \frac{(\gamma - 1)\eta}{\rho}\mathbf{j}\cdot\mathbf{j} = 0,
\end{equation}
where $p$ is the gas pressure, 
$\mathbf{B}=\nabla\times\left(\psi\hat{\mathbf{e}}_z\right) + \left(0,0,B_z\right)$ 
is the magnetic field ($\psi$ is the magnetic flux function),
$\mathbf{j}=\nabla\times\mathbf{B}$ is the current density,
$\eta$ is the resistivity,
and $\mathbf{F}$ is an additional force which is introduced in section \ref{subsection:extforce}.
The seven independent variables are the plasma density ($\rho$), temperature ($T$), velocity ($v_x$, $v_y$, $v_z$), 
magnetic flux function ($\psi$), and perpendicular component of magnetic field ($B_z$). 

The units of length, density and temperature are $L_0=1\times10^{9}$\ cm, $\rho_0=1.6\times 10^{-15}$\ g cm$^{-3}$ 
(i.e., $n_0=1\times 10^9$\ cm$^{-3}$), and $T_0=1\times 10^6$\ K, respectively. 
Velocity is normalized by
\begin{equation}
v_0 = C_s = \sqrt{\frac{2k_\mathrm{B}T_0}{m_\mathrm{H}}} = 128.6 \ \mathrm{km\ s}^{-1}
\end{equation}
where $k_\mathrm{B}$ is Boltzmann constant and $m_H$ is the mass of a hydrogen atom.

The parameter $\beta_0$ is the typical ratio of gas to magnetic pressures in the lower corona above an active region.
$\beta_0$ is chosen to be 0.01 measured at $(x,y)=(0,0.12)$  (magnetic configuration is described in section \ref{subsection:init}), and then the unit of magnetic field strength is
\begin{equation}
B_0 = \sqrt{\frac{16\pi\rho_0k_\mathrm{B}T_0}{m_\mathrm{H}\beta_0}} = 25.78\ \mathrm{G}.
\end{equation}
The unit of time is given by $\tau_{\rm A0} = L_0 / v_{\rm A0} $,
where $v_{\rm A0} $ is the Alfv\'{e}n velocity
($v_{\rm A0} = B_0 /\sqrt{4 \pi \rho_0} =1818$ km s$^{-1}$) 
and $\tau_{\rm A0} = 5.5 $ s. 
The resistivity ($\eta$) is described in section \ref{subsection:resist}.

The size of the simulation box is $-8 \le x \le 8$ in the horizontal direction and $0 \le y \le 15$ in the vertical direction.
The domain is divided by $201 \times 601$ grid points,
which are uniformly distributed in the $y$-direction while they are not distributed uniformly in the $x$-direction.
The grid spacing $\Delta x$ ranges from $6.46\times10^{-3}$ to $3.14\times10^{-1}$ with the finest interval around $x=0$.
At the bottom of the simulation box is applied a line-tying boundary condition where all quantities except $T$ 
are fixed; $T$ is determined by extrapolation.
At the other three boundaries is applied an open boundary condition.

\subsection{Initial Condition}\label{subsection:init}

\begin{figure}
\begin{center}
\plotone{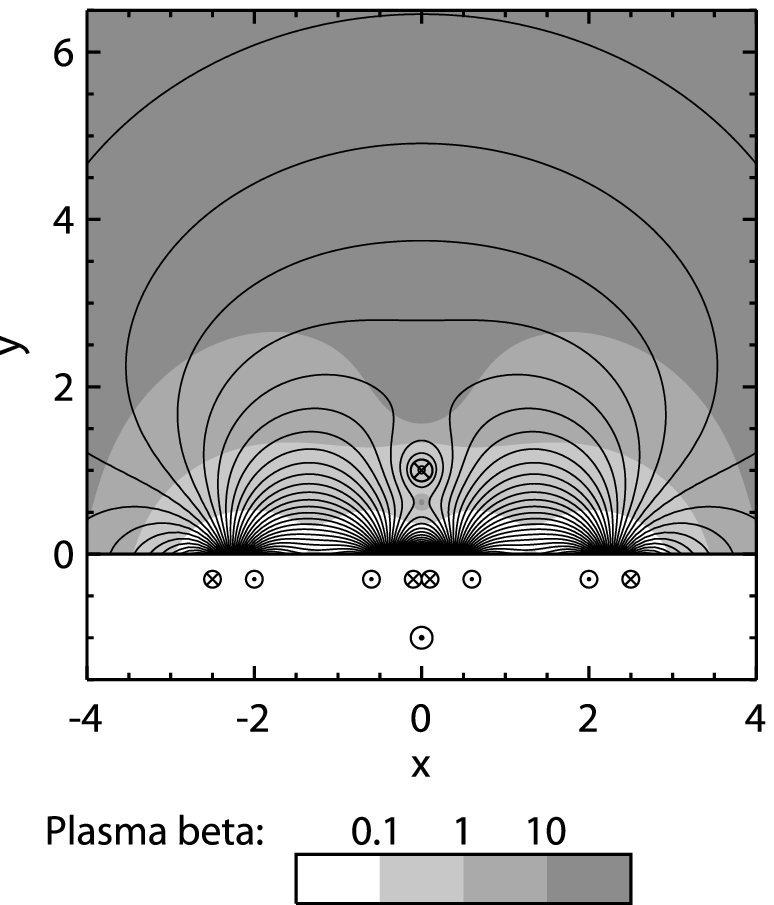}
\caption{A portion of the initial magnetic field configuration is presented.
The size of the simulation box is $-8 \le x \le 8$ in the horizontal direction and $0 \le y \le 15$ in the vertical direction.
Solid lines show magnetic field lines.
$\otimes$ and $\odot$ denote the direction of currents.
Gray scale shows the plasma beta, which is the ratio of gas pressure to magnetic pressure.}
\label{figure:initmagconfig}
\end{center}
\end{figure}

We assumed an octapolar magnetic configuration as the initial condition (Figure \ref{figure:initmagconfig}). 
Our model is based on the model of \citet{2000ApJ...545..524C} and \citet{2005ApJ...634..663S} that assumes a quadrupolar configuration.
We added two dipole fields to this model, in order to control the magnetic flux in the inflow region, duration and released energy of a flare.
Short durations are the feature of impulsive flares.

To get the magnetic configuration with a detached plasmoid (or a flux rope in 3D situation) and two side loops, our initial state consists of three separate groups of currents in a two-dimensional Cartesian plane: line current placed at the point (0, $h$) with finite radius $r_0$, its image current below the photosphere, and a background octapolar potential field produced by eight line currents below the photosphere, i.e., at eight positions $(\pm 0.1, -0.3)$, $(\pm 0.6, -0.3)$, $(\pm 2.0, -0.3)$, and $(\pm 2.5, -0.3)$, with the direction being ``$+$, positive $z$'', ``$-$, negative $z$'', ``$-$', ``$+$'', ``$+$'', ``$-$'', ``$-$'', and ``$+$'' in sequence.
Then, the initial magnetic configuration is expressed as 
\begin{equation}
\psi = \psi_b + \psi_i + \psi_l,
\end{equation}
where the background field ($\psi_b$) and the magnetic components of the image current ($\psi_i$) and of the line current ($\psi_l$) have the following forms:

\begin{eqnarray}
\psi_b &=& c_1 \ln
\left( \frac{[(x+0.1)^2+(y+0.3)^2][(x-0.1)^2+(y+0.3)^2]}{[(x+0.6)^2+(y+0.3)^2][(x-0.6)^2+(y+0.3)^2]} \right) \nonumber \\
&+& c_2 \ln \left( \frac{[(x+2.5)^2+(y+0.3)^2][(x-2.5)^2+(y+0.3)^2]}{[(x+2.0)^2+(y+0.3)^2][(x-2.0)^2+(y+0.3)^2]} \right) ,\label{eqn:background} \\
\psi_i &=& - c_3 \frac{r_0}{2}\ln[x^2+(y+h)^2], \label{eqn:image} \\
\psi_l &=& \left\{
\begin{array}{ll}
\displaystyle c_4 \frac{r^2}{2r_0}, &  (r \leq r_0), \\
c_4 \left[ r_0/2 - r_0 \ln(r_0) + r_0 \ln(r)  \right], &  (r > r_0),
\end{array}
\right. \label{eqn:rope}
\end{eqnarray}
where $r=\sqrt{x^2+(y-h)^2}$ is the distance from the center of the plasmoid,
$h$ is the height of the plasmoid and $r_0$ is the radius of the plasmoid,
which determine the configuration of magnetic field.
In this study, we set $h=1.0$ and $r_0=0.2$.

The coefficients $c_1$, $c_2$, $c_3$ and $c_4$ in the equations (\ref{eqn:background}), (\ref{eqn:image}) and (\ref{eqn:rope}) control the strength of the magnetic field.
When the radius of the plasmoid, $r_0$, is small enough compared to the height of the plasmoid,
$h$, equilibrium solutions are analytically given by \citet{2005ApJ...629..582L}.
However, it is difficult to determine the coefficients analytically in our model,
because the plasmoid is in the complex background magnetic field and $r_0$ is not too small.
Therefore we determine the coefficients, $c_1$, $c_2$, $c_3$, and $c_4$, by trial and error
in order to make the plasmoid center keep roughly stable and move much slowly compared to a rapid motion driven by reconnection.
In this study, we set the coefficients as $c_1=2$, $c_2=2$, $c_3=6$ and $c_4=2.4$.

The physical meaning is as follows.
If we remove the plasmoid, the magnetic field configuration  
are potential octapolar field, produced by the eight line currents, $\psi_b$,
and the image current below the photosphere, $\psi_i$.
The configuration makes a null point, which is saddle of potential above the photosphere.
If we place a plasmoid with zero radius at the null point, then the plasmoid does not begin to move itself without any perturbation.

Although it is difficult to create a state of equilibrium when a plasmoid has a finite radius, this is not a major problem because the purpose of this study is not to investigate the trigger mechanism of a flare, but to focus on the physical processes working after magnetic reconnection starts.

To satisfy the force balance within the plasmoid,
a perpendicular magnetic component (i.e., $B_z$) is introduced inside the plasmoid:
\begin{equation}
B_z=\left\{ \begin{array}{ll}
 0.2\sqrt{2\left(1-{\displaystyle \frac{r^2}{r_0^2}}\right)}, & r \leq r_0 ;\\
 0, & r > r_0,
        \end{array}
  \right.
\end{equation}
Other quantities are set to be uniform; 
$(\rho, T, v_x, v_y, v_z)=(1.0, 1.0, 0.0, 0.0, 0.0)$.

\subsection{Resistivity}\label{subsection:resist}

The resistivity, $\eta$, is assumed as an anomalous resistivity:
\begin{equation}
\eta=\left\{ \begin{array}{ll}
\eta_0, &
v_{\rm d} \geq 2v_{\rm c},\\
\eta_0\left({\displaystyle \frac{v_{\rm d}}{v_{\rm c}}}-1\right), &
v_{\rm c} \leq v_{\rm d} < 2v_{\rm c},\\
0, & v_{\rm d} < v_{\rm c} ,
\end{array}
\right. \label{eqn:resist}
\end{equation}
where $v_{\rm d} \equiv {|j_z|}/{\rho}$ is the (relative ion-electron) drift velocity, and $v_{\rm c}$ is the critical velocity.
Here we assumed $v_c=0.5$.
The anomalous resistivity begins to work when the drift velocity exceeds the critical velocity, otherwise there is no resistivity, only small numerical resistivity remains.
It is known that the anomalous resistivity may be caused by plasma instabilities \citep{1971ApJ...169..379C},
and localized resistivity triggers fast reconnection \citep{1977JPlPh..17..337U}.
We assumed that $\eta_0$ is a free parameter (see Table \ref{table:resisttable}).
We simulated with various $\eta_0$ in order to change the reconnection rate (case A).
Therefore we can examine how the reconnection rate influences the plasmoid velocity.
We also set the resistivity to be zero in order to prevent magnetic reconnection and investigated the role of reconnection in plasmoid ejection (case C).

\begin{deluxetable}{ccc}
\tablecaption{Free parameter $\eta_0$ which determines the resistivity and strength of the additional force ($F_y$).\label{table:resisttable}}
\tablehead{\colhead{Case} & \colhead{$F_y$} & \colhead{$\eta_0$}}
\startdata
 A   &  0 &  0.2, 0.02, 0.002, 0.0002\\
 B   & $\pm0.5$, $\pm1.0$ &  0.002  \\
 C   &  0 &  0 \\
\enddata
\end{deluxetable}

\subsection{Additional Force} \label{subsection:extforce}

Considering the plasmoid-induced-reconnection model \citep{1996AdSpR..17....9S, 1997cswn.conf..103S}, 
we can expect that there is a positive correlation between the plasmoid velocity and reconnection rate.
To see this, we changed the vertical velocity of the plasmoid by using the vertical additional force that acts only inside the plasmoid (or a flux rope in a 3D situation) in case B.
There is no additional force in case A and case C.
The additional force is introduced as a virtual force in order to change the plasmoid velocity and see how it relates to the reconnection rate.
\footnote{The additional force used in the present simulations is a purely virtual force, and we do not intend to model the force realistically. For the interested reader, however, it may be useful to note that there have been proposed several ways of modeling the force in a 3D space, such as curvature force (see \citealt{1966RvPP....2..103S} and \citealt{1998ApJ...504.1006L}).}
In the present study we do not focus on the role of the additional force in triggering a flare,
rather we investigate how the force affects the evolution following the onset of a flare.
In fact, even if there is no additional force applied to the initial state,
the plasmoid begins to move upward due to the non-equilibrium of the initial state.
The additional force $\mathbf{F}$ in equation (\ref{fqn:motion}) is defined as follows:
\begin{equation}
\mathbf{F}=\left\{ \begin{array}{ll}
 (0, F_y F_0, 0), & r \leq r_0 ;\\
 (0, 0, 0),       & r > r_0,
        \end{array}
  \right.
\end{equation}
where $F_0 = 0.271 \rho_0 L_0 / C_s$, and $F_y$ is given in Table \ref{table:resisttable}.
Note that the value of $F_y$ is also set to be negative, which means that the direction of additional force is downward and it decelerates the plasmoid.

\section{Results}\label{section:results}

\begin{figure}[p]
\begin{center}
\epsscale{0.9}
\plotone{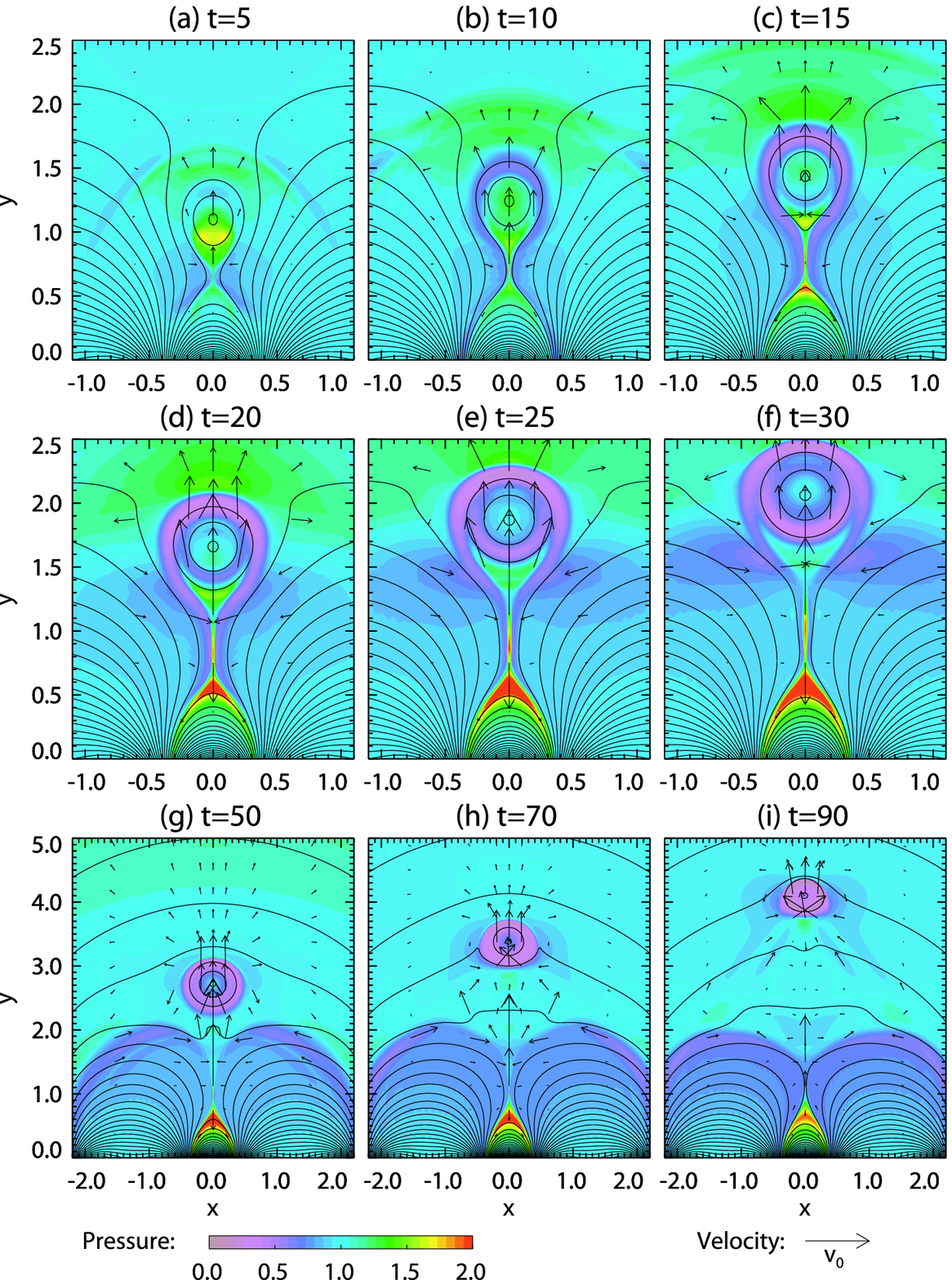}
\caption{Pressure distribution in a typical case (case A with $\eta_0=0.002$). 
The solid lines denote magnetic field lines, and arrows represent velocity field.}
\label{figure:typicalcase_pressure}
\end{center}
\end{figure}

\begin{figure}[p]
\begin{center}
\epsscale{1.0}
\plotone{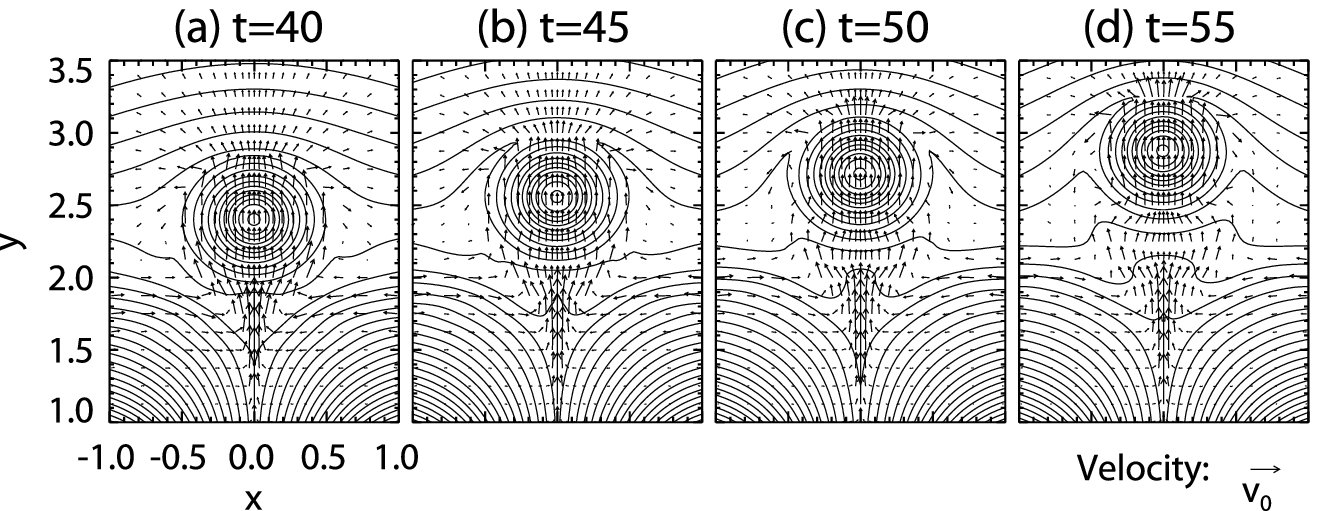}
\caption{The magnetic field and velocity field nearby the plasmoid in case A, $\eta_0=0.002$. The solid lines denote magnetic field lines, and arrows represent velocity field. Magnetic reconnection occurs at the top of the plasmoid.}
\label{figure:plasmoid}
\end{center}
\end{figure}

\begin{figure}[p]
\begin{center}
\plotone{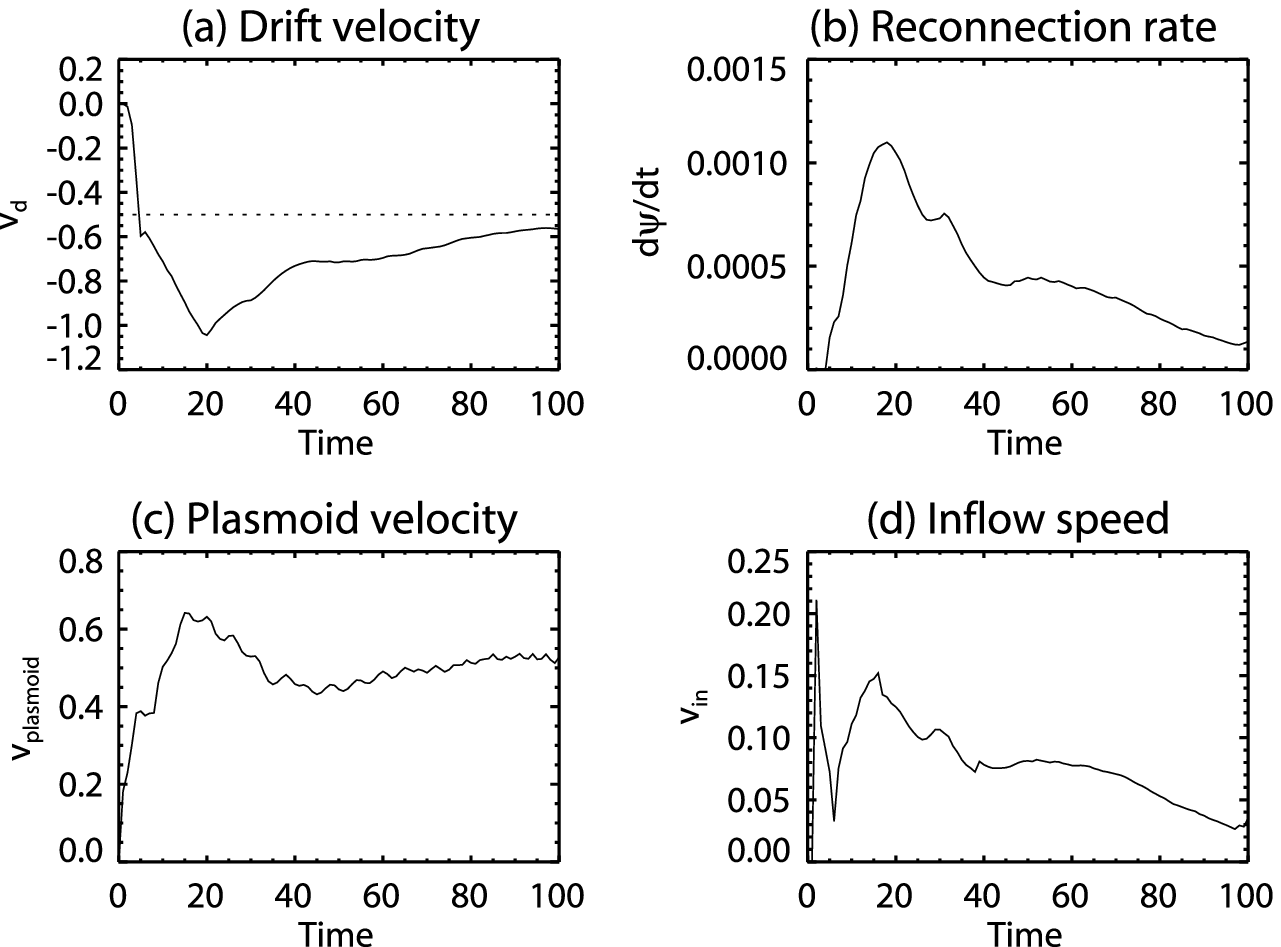}
\caption{These panels show the temporal variation of the standard case (case A, $\eta_0=0.002$).
(a) Maximum drift velocity of $v_d=j_z/\rho$ in the current sheet. Dotted line shows $-v_c$.
(b) Maximum reconnection rate in the current sheet, not normalized (i.e. $d\psi/dt = \eta j$).
(c) Plasmoid velocity.
(d) Inflow speed measured at the point of maximum reconnection rate.}
\label{figure:typicalcase_graph}
\end{center}
\end{figure}

\begin{figure}[p]
\begin{center}
\plotone{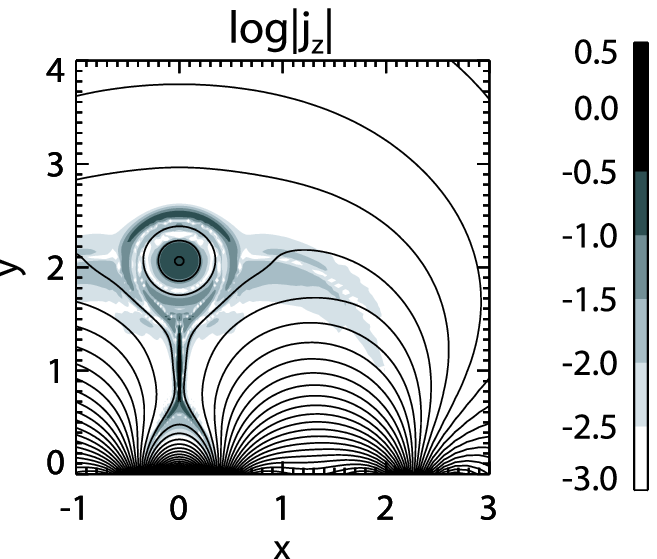}
\caption{Distribution of $z$-component of current density ($j_z$) at $t=30$ in the standard case (case A, $\eta_0=0.002$). 
The solid lines denote magnetic field lines.}
\label{figure:typicalcase_current}
\end{center}
\end{figure}

\begin{figure}[p]
\begin{center}
\plotone{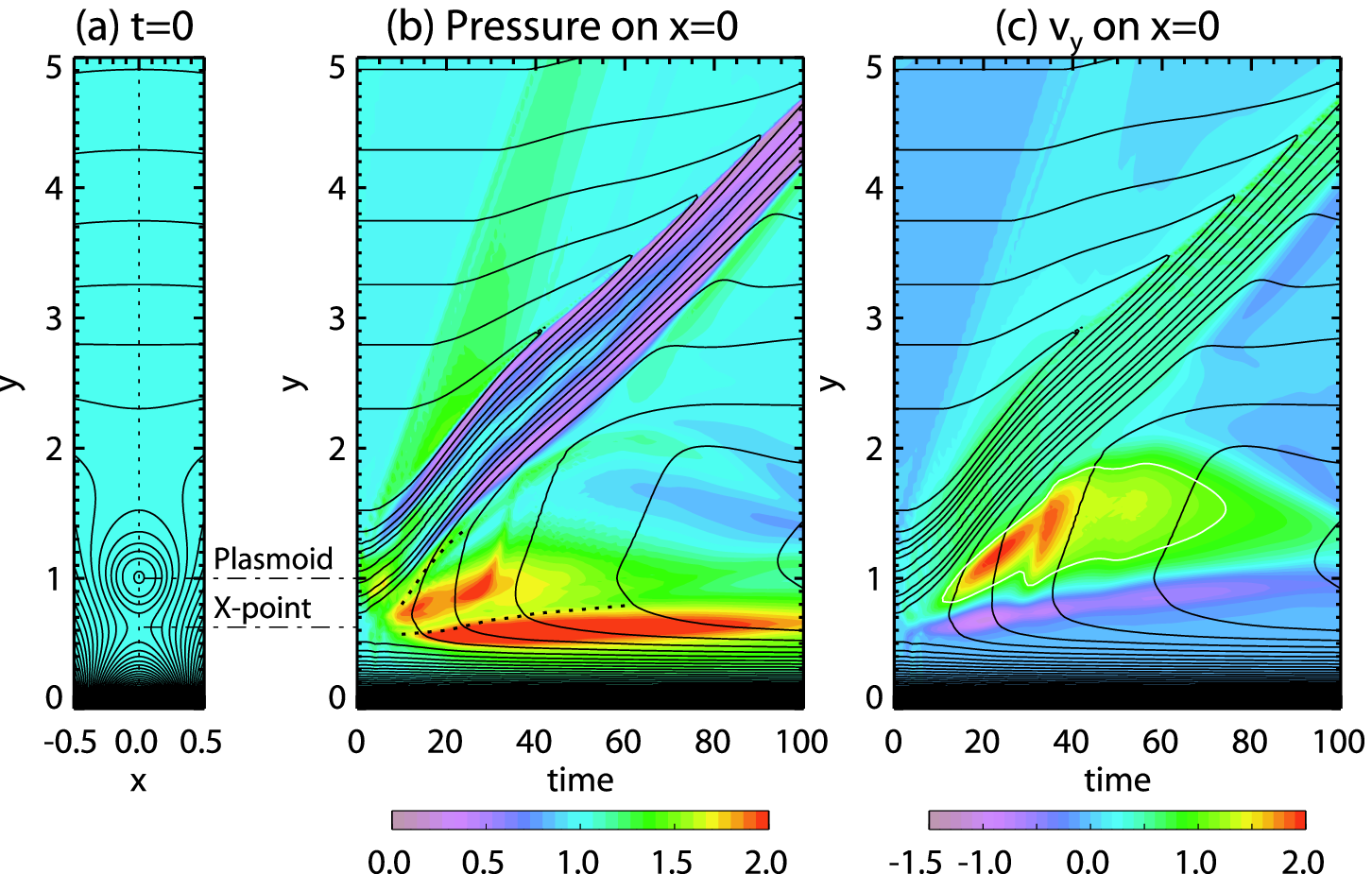}
\caption{Panel (a) shows the initial configuration of magnetic field, in which solid lines represent the distribution of magnetic flux $\psi$. At panels (b) and (c) the contours show the time evolution of the point on each field line that is located at the $y$-axis in the standard case (case A, $\eta_0=0.002$). The colors variation show the temporal change of gas pressure in panel (b) and vertical velocity $v_y$ in panel (c). The dotted lines in panel (b) show the trajectory of the fast shocks (termination shocks) formed at the top of the flare loop and the bottom of the plasmoid. The white line in panel (c) shows the position of the upward jet where $v_y>v_0$.}
\label{figure:timeslice}
\end{center}
\end{figure}

\begin{figure}
\begin{center}
\epsscale{0.5}
\plotone{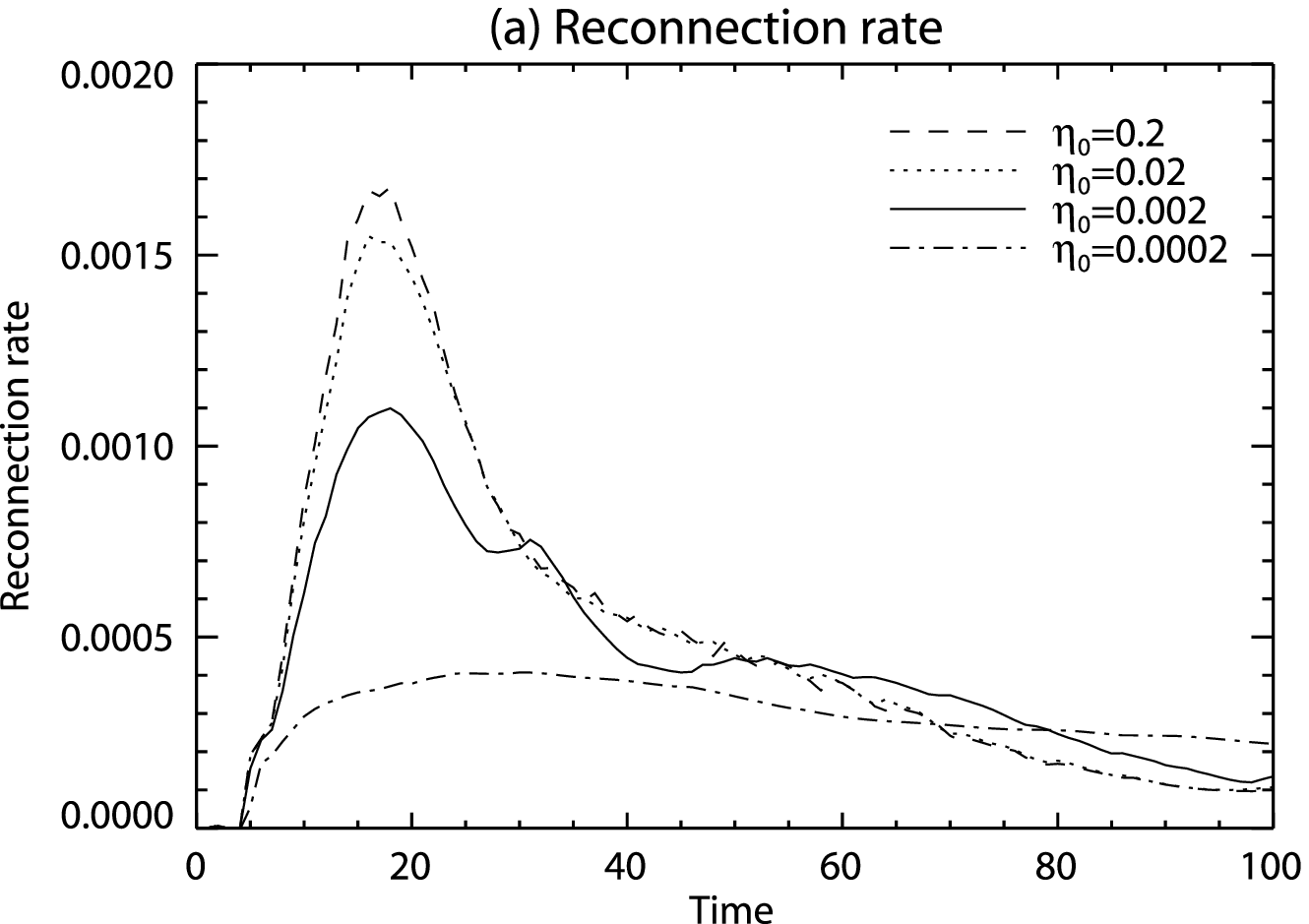}
\plotone{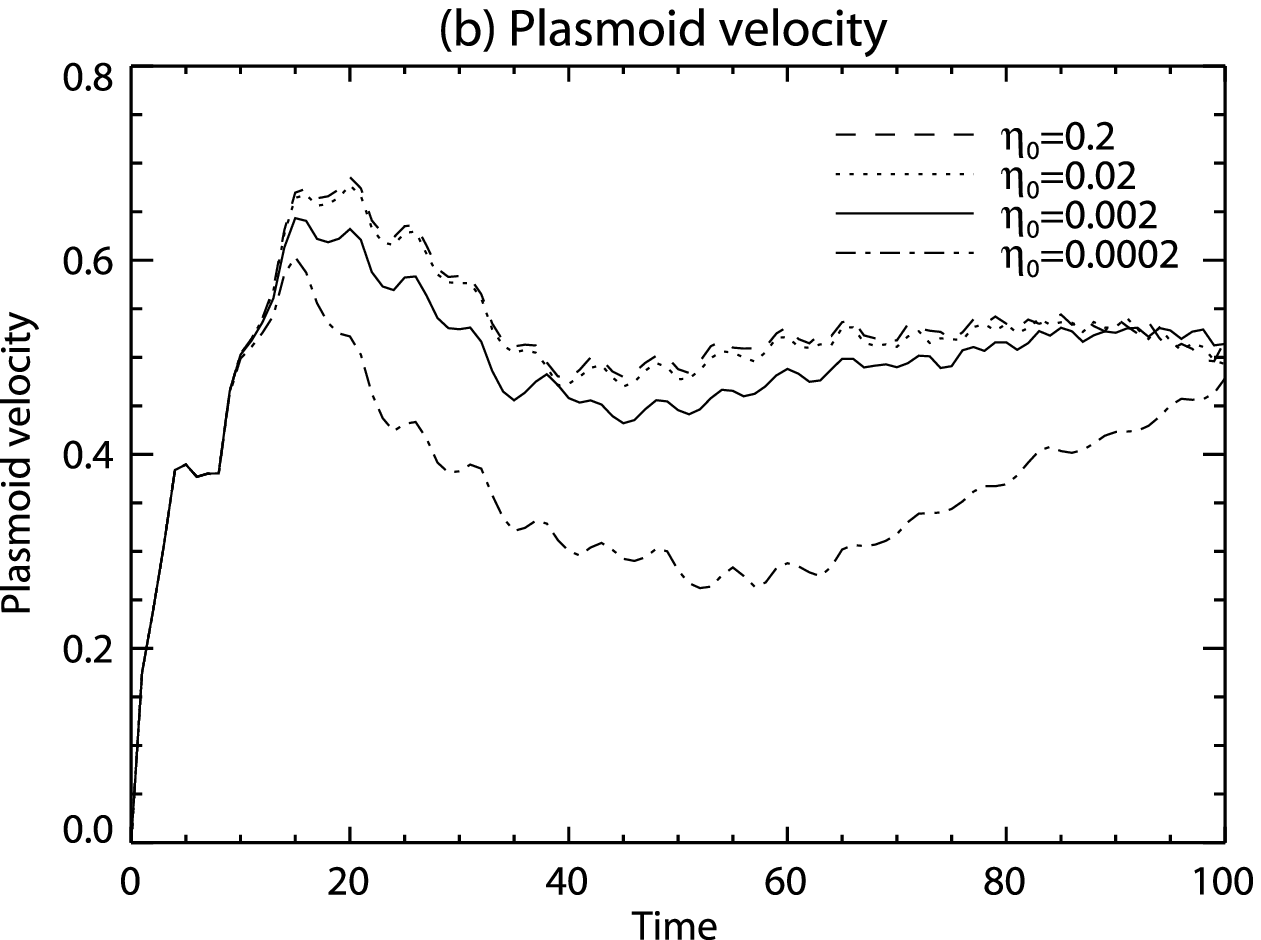}
\plotone{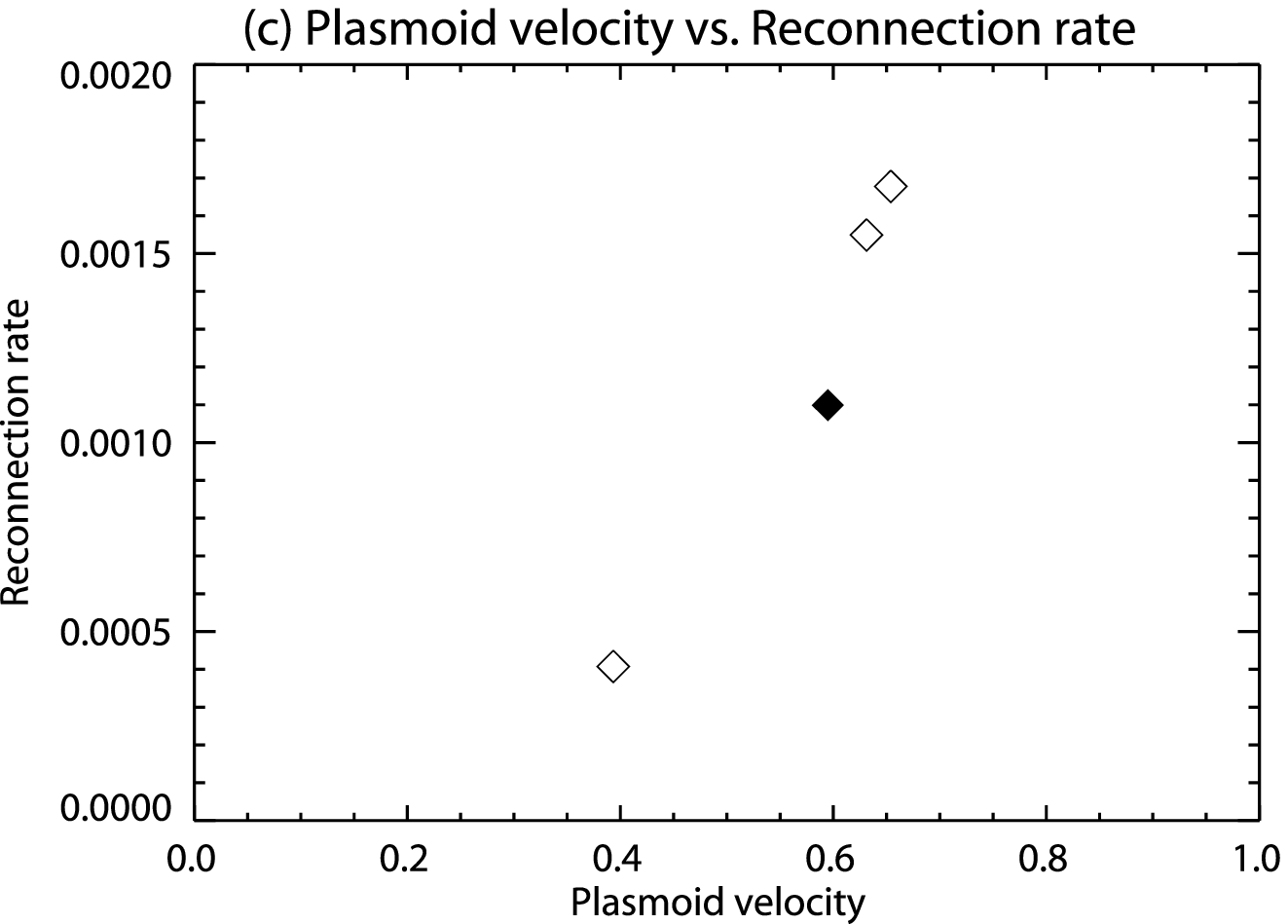}
%\plotone{pf_lr3_vc05.eps}
%\plotone{vfy_lr3_vc05_20080504.eps}
\caption{Time variation of (a) the reconnection rate ($d\psi/dt$) and (b) the plasmoid velocity ($v_{plasmoid}$) in case A, when free parameter $\eta_0$ is changed. Panel (c) shows the correlation between the maximum value of the reconnection rate and the plasmoid velocity when the reconnection rate becomes the maximum in case A. The black diamond is the standard case, $\eta_0=0.0002$.}
\label{figure:lr3_vc05}
\end{center}
\end{figure}

\begin{figure}
\epsscale{1.0}
\plottwo{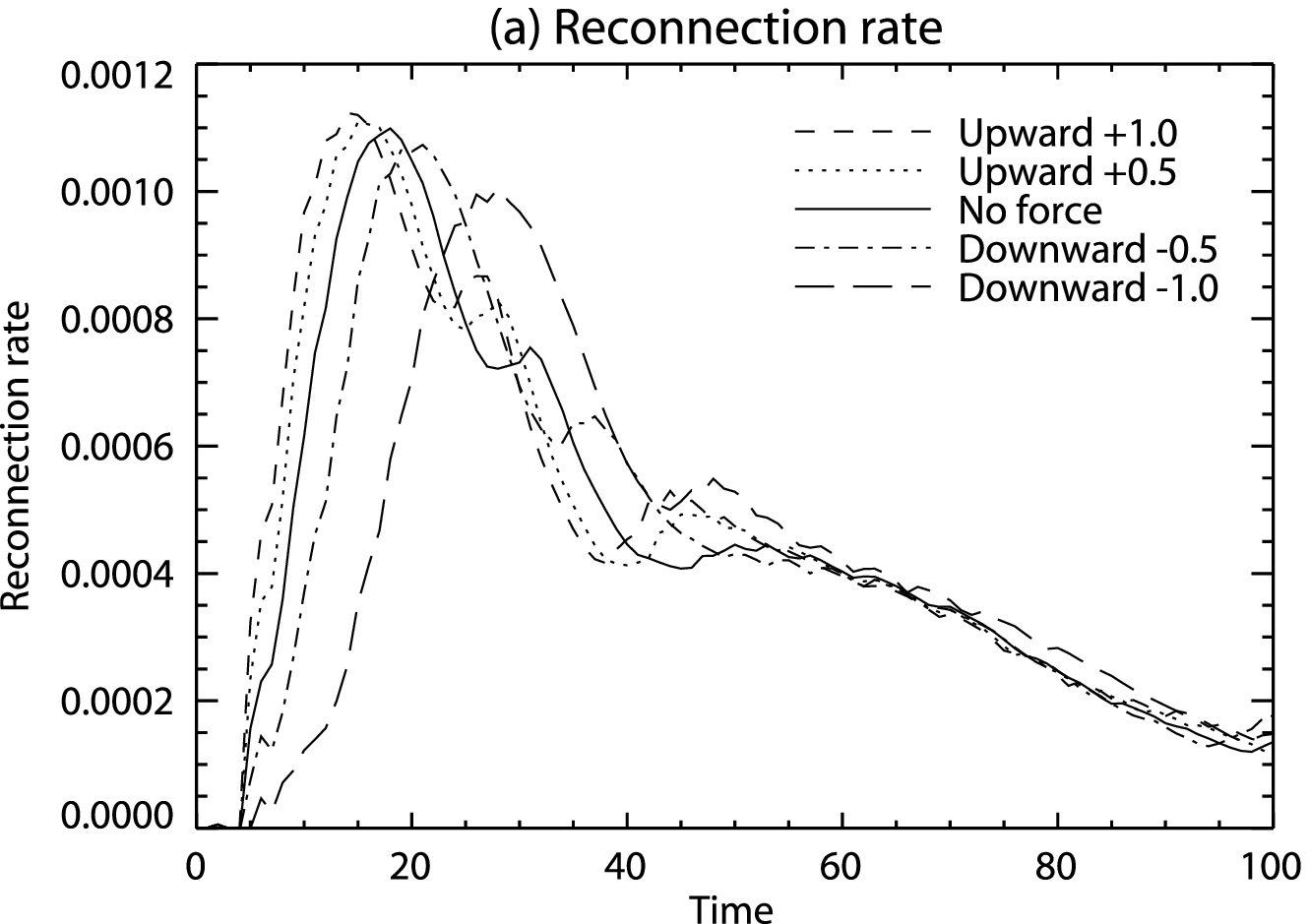}{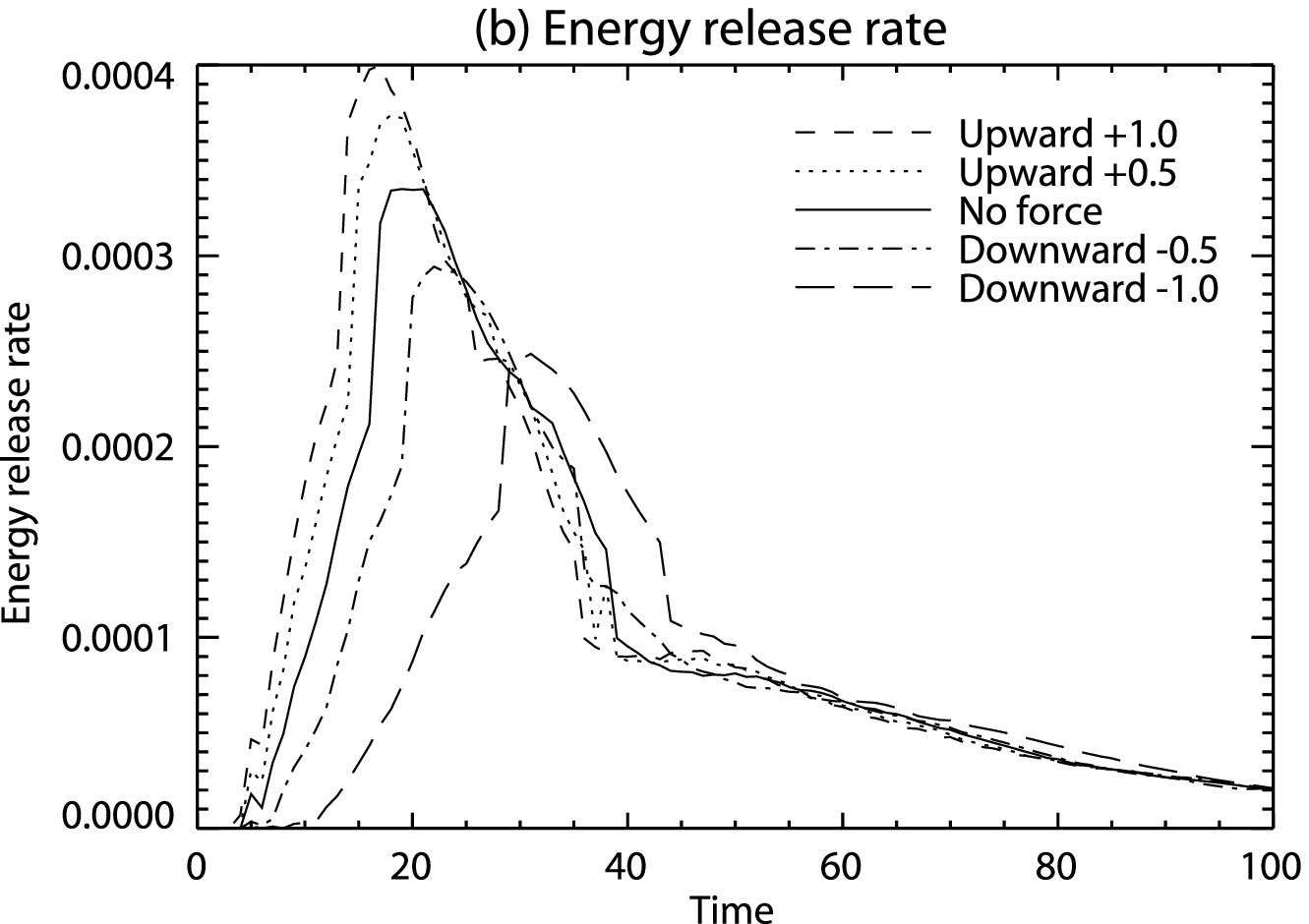}
\plottwo{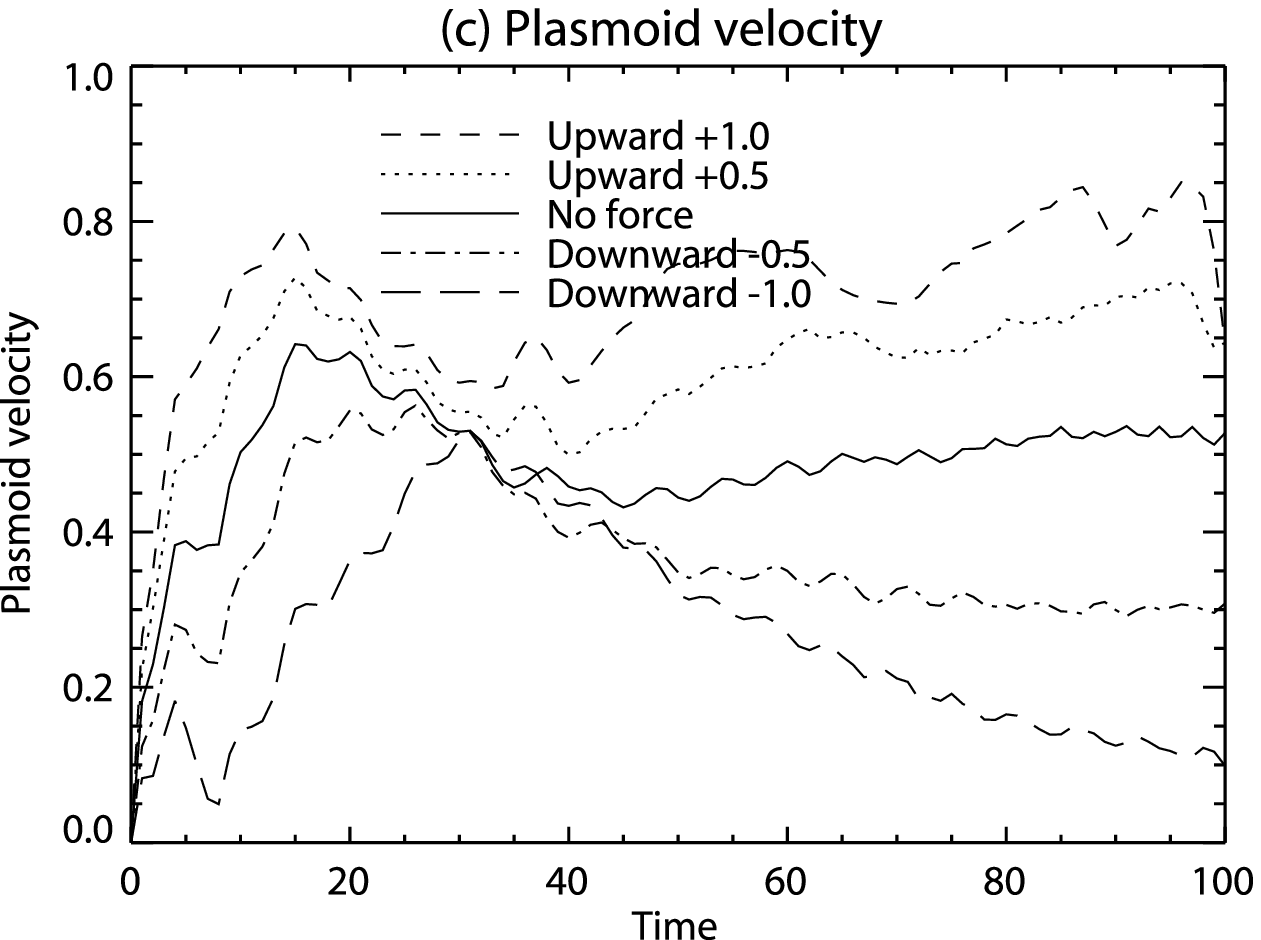}{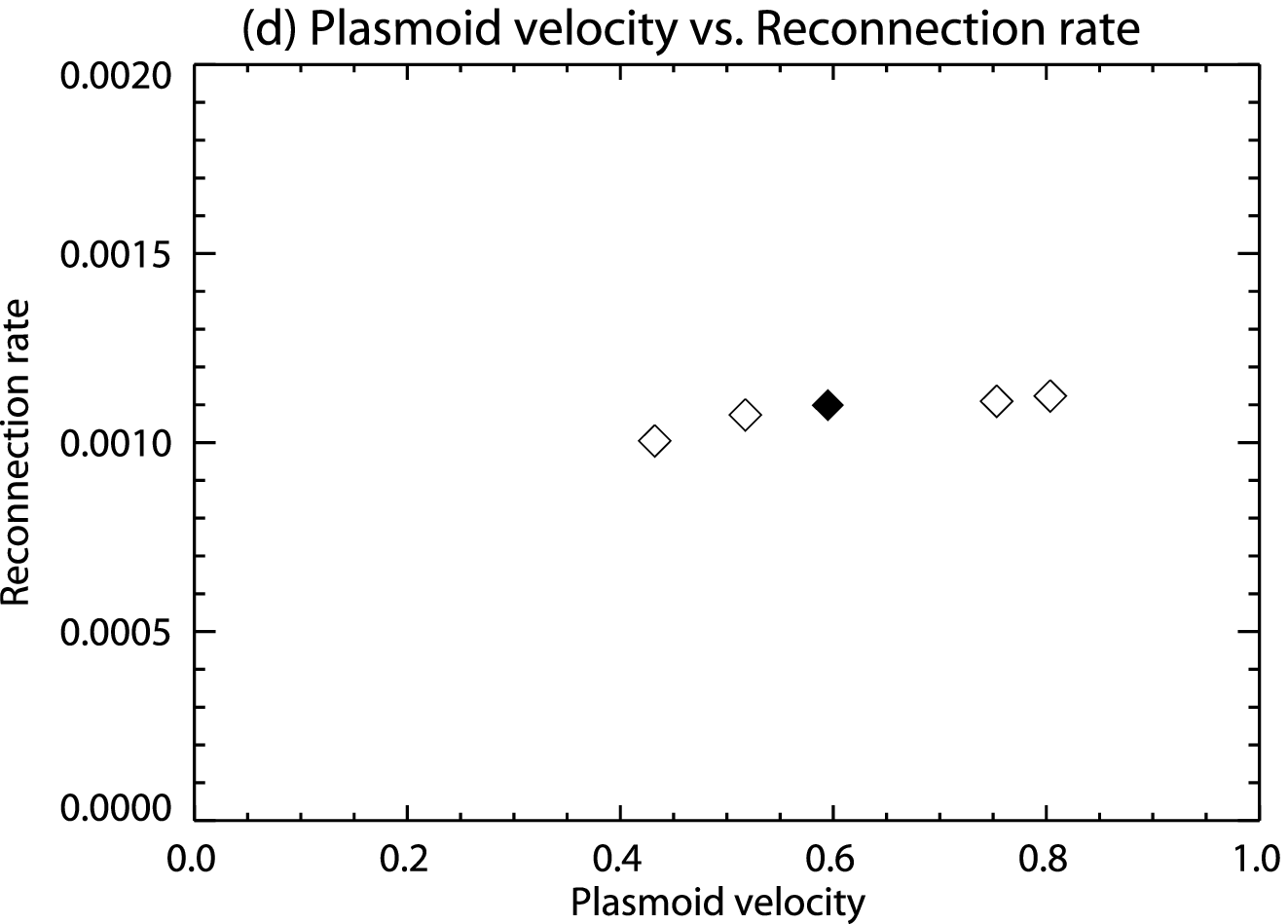}
\caption{Time variation of (a) the reconnection rate ($d\psi/dt$), (b) the energy release rate, and (c) the plasmoid velocity ($v_{plasmoid}$) in case B, when the additional force $F_y$ is changed. Panel (d) shows the correlation between the maximum value of the reconnection rate and the plasmoid velocity when the reconnection rate becomes the maximum. The white diamonds show the result in in case B. Solid lines in panel (a)(b)(c) and the black diamond in panel (d) show the result in standard case (case A, $\eta_0=0.002$), just for comparison.}
\label{figure:ef3_vc05}
\end{figure}

In this section, we introduce the typical result from case A.
Figures \ref{figure:typicalcase_pressure} and \ref{figure:plasmoid} show the magnetic field distribution on $x$-$y$ plane and Figure \ref{figure:typicalcase_graph} shows the temporal variation of physical values in case A with $\eta_0=0.002$, which is called the standard case here after.
Because of the non-equilibrium of the initial state, the plasmoid begins to move upward a little in $0<t<5$ in Figure \ref{figure:typicalcase_graph}c, \ref{figure:timeslice}, \ref{figure:lr3_vc05}b, \ref{figure:ef3_vc05}c  (i.e. melon-seed effect).
After that the plasmoid starts to be accelerated when the reconnection starts until $t \sim 15$.
The rising plasmoid causes the inflow of magnetized plasma into the magnetic null point (X-point).
Magnetic field lines at both sides of the null point are carried into the point, and a current sheet is formed near the X-point. 
The drift velocity inside the current sheet then increases quickly by the inflow in $2<t<5$ (Figure \ref{figure:typicalcase_graph}a).
When the condition of equation (\ref{eqn:resist}) is satisfied, anomalous resistivity sets in and fast reconnection occurs (Figure \ref{figure:typicalcase_graph}b).
This accompanies Y-shape and inverted Y-shape slow shocks at both ends of the current sheet (Figure \ref{figure:typicalcase_current}), which is one of the features of Petschek-type reconnection \citep{1964psf..conf..425P}.
The plasmoid is accelerated rapidly in $5<t<15$ (Figure \ref{figure:typicalcase_graph}c).
This result is consistent with observations \citep{1997PASJ...49..249O,1998ApJ...499..934O,2004ApJ...613..592T,2001ApJ...559..452Z,2005ApJ...630.1148S} and a simulation \citep{1997ApJ...487..437M}.
However the plasmoid is decelerated gradually during $15 < t < 40$, although
reconnection jets push the bottom of the plasmoid (the white line in Figure \ref{figure:timeslice}c).
This deceleration is mainly due to gas pressure, since the plasma beta above the plasmoid is high (see plasma beta distribution in Figure \ref{figure:initmagconfig}).
The rising plasmoid induces a strong inflow into the X-point in $5<t<40$ (Figure \ref{figure:typicalcase_graph}d), and consequently the reconnection rate grows up.
Here we define the reconnection rate as flux canceling speed i.e. $d\psi/dt (=\eta j)$ at the X-point.

Figure \ref{figure:timeslice} shows the time evolution of pressure and vertical velocity at $x=0$.
The bipolar jets from X-point are created by reconnection.
The velocity of the upward jet reaches up to the local Alfv\'{e}n velocity.
The jets created a fast-mode shock when colliding with the magnetic loop and the plasmoid.
The upper fast shock is weaker than the lower one, because the relative speed between the reconnection jet and preexisted plasma is smaller.
The lower fast-mode shock (termination fast shock) may be observed as a loop top hard X-ray (HXR) source \citep{1994Natur.371..495M}.
The termination fast shock is rising as the magnetic loop develops due to the piling-up of reconnected field.

Figure \ref{figure:lr3_vc05} shows time variation of the reconnection rate and the plasmoid velocity in case A with different values of $\eta_0$.
The reconnection rate and the plasmoid velocity are enhanced as $\eta_0$ grows.
The reconnection rate seems to be proportional to $\log(\eta)$ except $\eta_0=0.2$,
but a careful analysis with wider parameter survey is needed for a final conclusion.

The time variations of the reconnection rate, the energy release rate, and the plasmoid velocity are shown in Figure \ref{figure:ef3_vc05}, when we changed the plasmoid velocity due to the additional force (case B).
An interesting point suggested by this figure is that the plasmoid velocity changes significantly when the additional force is applied, but the reconnection rate does not.
On the other hand, the energy release rate ($B_{inflow}^2/4\pi \times v_{inflow} L_{inflow}$ near the current sheet) is changed significantly, because the length of the current sheet ($L_{inflow}$) is changed by the position of the plasmoid.

Figure \ref{figure:lr3_vc05}c and Figure \ref{figure:ef3_vc05}d show the correlation between the reconnection rate and the plasmoid velocity measured when the reconnection rate becomes the maximum.
A positive correlation is found between the reconnection rate and the plasmoid velocity in case A (Figure \ref{figure:lr3_vc05}c).
But in case B, there is no strong correlation between the reconnection rate and the plasmoid velocity (Figure \ref{figure:ef3_vc05}d).

\begin{figure}
\epsscale{1.0}
\plotone{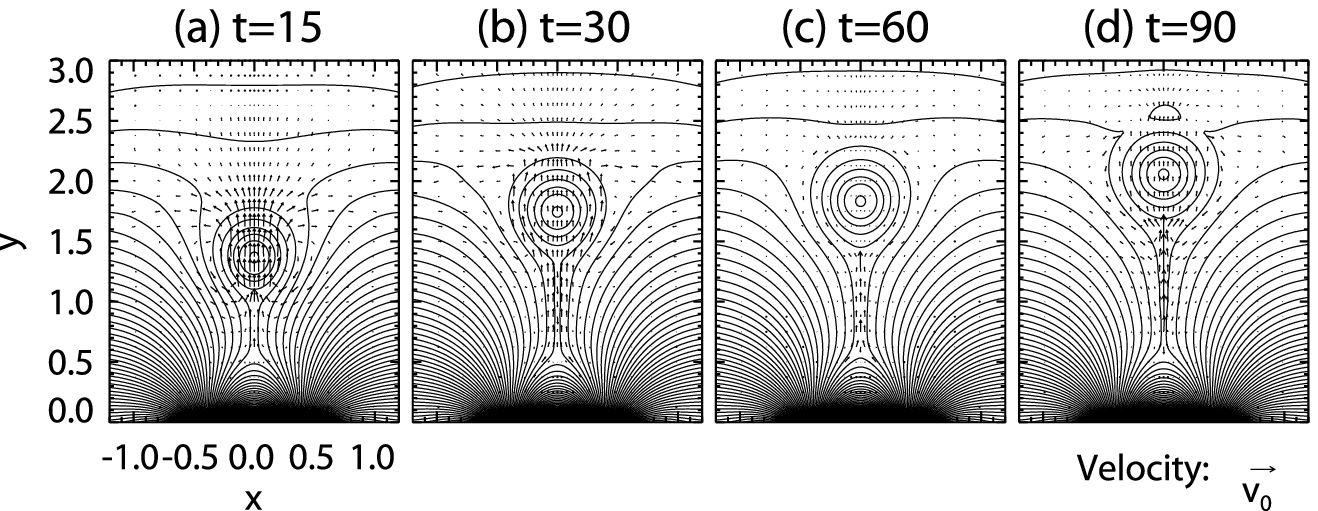}
\caption{The magnetic field and velocity field in the case of no resistivity (case C).
The solid lines denote magnetic field lines, and arrows represent velocity field.}
\label{figure:noresist}
\end{figure}

Figure \ref{figure:noresist} shows the magnetic field and velocity field in the case of no resistivity (case C, $\eta_0=0$).
The plasmoid begins to move upward slightly without reconnection, due to the non-equilibrium of the initial state (Figure \ref{figure:noresist}a).
The rising plasmoid pulls up the magnetic field and stretches the current sheet,
but reconnection does not start due to the lack of the resistivity.
Therefore the plasmoid is still trapped by the closed magnetic field and stops rising (Figure \ref{figure:noresist}c).
This result shows the importance of reconnection in plasmoid ejection.
In fact, the reconnection based on numerical resistivity occurred in a later time, so the plasmoid began to move upward again (Figure \ref{figure:noresist}d).

\section{Discussion}\label{section:discussion}

Now we shall discuss the comparison of the simulation results with observation.
From soft X-ray observations of plasmoid ejections in compact impulsive flares,
\citet{1995ApJ...451L..83S} and \citet{shimizu2008} found that there is a positive correlation between the plasmoid velocity and the rise velocity of the post-flare loop.
\citet{2005ApJ...634L.121Q} also found a similar relation between the speed of coronal mass ejections (CMEs) and the reconnection rate.
The rise velocity of the post-flare loop is proportional to the reconnection inflow speed,
so this means that there is a positive correlation between the plasmoid velocity and the reconnection inflow speed.
\citet{2006ApJ...637.1122N} discussed a correlation between CMEs speed and the inflow speed.
This is actually obtained by the present result.

How is this result explained physically?
In case A, when the resistivity increases, the reconnection rate ($\eta J$) increases.
Since $\eta J = v_{in} B$ in steady state, this means that the inflow speed ($v_{in}$) increases.
Accordingly, the reconnection jet speed ($v_{out}$) also increases when $v_{in}$ increases
because the mass conservations hold, $v_{in} L_{in} = v_{out} W_{out}$,
in the incompressible limit.

\begin{figure}
\plotone{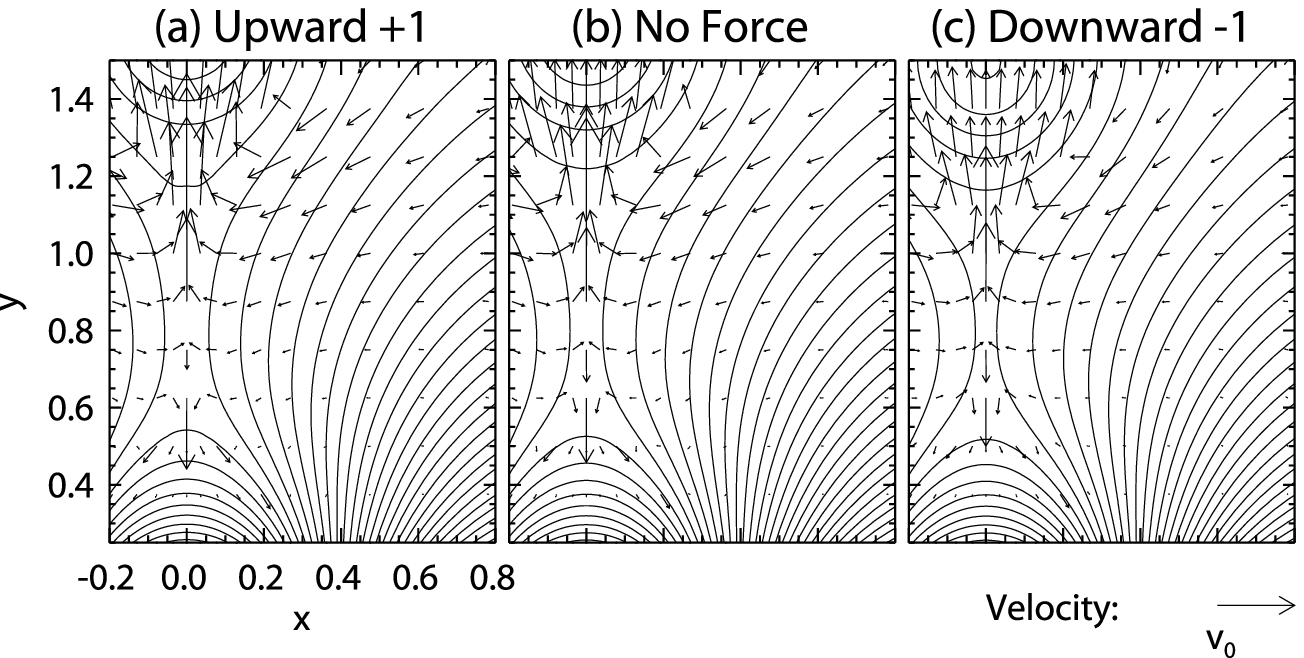}
\caption{The magnetic field and velocity field nearby the inflow region in case B, when the additional force $F_y$ is changed.
The solid lines denote magnetic field lines, and arrows represent velocity field.
The times are $t=14, 18, 28$ respectively, when the position of the plasmoid is $y=1.5$ approximately.
}
\label{figure:inflowvelocity}
\end{figure}

We also presented a physical explanation of how the plasmoid velocity is related to the enhancement of the resistivity in case B.
In this case, we changed the plasmoid velocity by applying the additional force.
We found weak dependence of the reconnection rate on the plasmoid velocity,
on the contrary to the fact that the reconnection rate is supposed to be proportion to the plasmoid velocity because of mass conservation law.
When the plasmoid velocity is larger, the plasmoid is accelerated farther from the X-point.
When the plasmoid is situated far from the X-point,
the plasmoid motion induces inflows only into the slow shock region near the plasmoid, but not into the X-point (Figure \ref{figure:inflowvelocity}).
The dependence of reconnection rate on the plasmoid velocity is small.
However the length of the slow shock increases when the plasmoid velocity is larger, so that energy release rate increases (Figure \ref{figure:ef3_vc05}b).

On the other hand there is a possibility that the positive correlation between the plasmoid velocity and the reconnection inflow speed holds in the initial phase, during which inflows will directly come into X-point.
To examine the initial phase is however difficult in our model, because the initial configuration of our model is not in equilibrium and we cannot measure small changes of the reconnection rate and inflow speed in the initial phase.
We will try to perform simulations with the initial condition in equilibrium in future.

It should be mentioned that in our model (Figure \ref{figure:initmagconfig}), magnetic flux stored in the plasmoid is lost by the magnetic reconnection between the magnetic field overlying the plasmoid and magnetic flux stored in the plasmoid \ref{figure:plasmoid}.
Therefore, if the amount of the magnetic flux stored in the plasmoid is small, the plasmoid disappears and does not propagate to the higher corona.
There is another role of the reconnection occurring above an ejecting plasmoid. If the reconnection does not occur at the top of the plasmoid, the plasmoid might be decelerated by magnetic tension force, while if the reconnection occurs, the plasmoid is not decelerated. However in this simulation the plasma beta is high above the plasmoid (see Figure \ref{figure:initmagconfig}), so the plasmoid is decelerated by gas pressure even if the reconnection occurs.

\section{Conclusion}\label{section:conclusion}

In this paper, we performed several MHD simulations to examine the basic physical relation between the plasmoid velocity and reconnection rate in the context of the plasmoid-induced-reconnection model for solar flares.
The initial magnetic configuration, resistivity, and the additional force are parameters in these simulations.
When we changed the amplitude of resistivity (case A), the reconnection rate and the plasmoid velocity changed,
showing a positive correlation.
We showed that the reconnection rate (i.e. inflow speed) and the plasmoid velocity are closely related to each other.
This result is consistent with observations \citep{1995ApJ...451L..83S, 2005ApJ...634L.121Q, shimizu2008} supporting the plasmoid-induced-reconnection model of impulsive flares.

\acknowledgments
Numerical computations were carried out in part on general common use computer system at the Center for Computational Astrophysics, CfCA, of the National Astronomical Observatory of Japan.
This work was supported by the Grant-in-Aid for the global COE program ``The Next Generation of Physics, Spun from Universality and Emergence'' from the Ministry of Education, Culture, Sports, Science and Technology (MEXT) of Japan and by the Grant-in-Aid for Creative Scientific Research of the MEXT ``The Basic Study of Space Weather Prediction'' (17GS0208, PI: K. Shibata).

\appendix

\section{An acceleration mechanism of a plasmoid}
\begin{figure}
\plotone{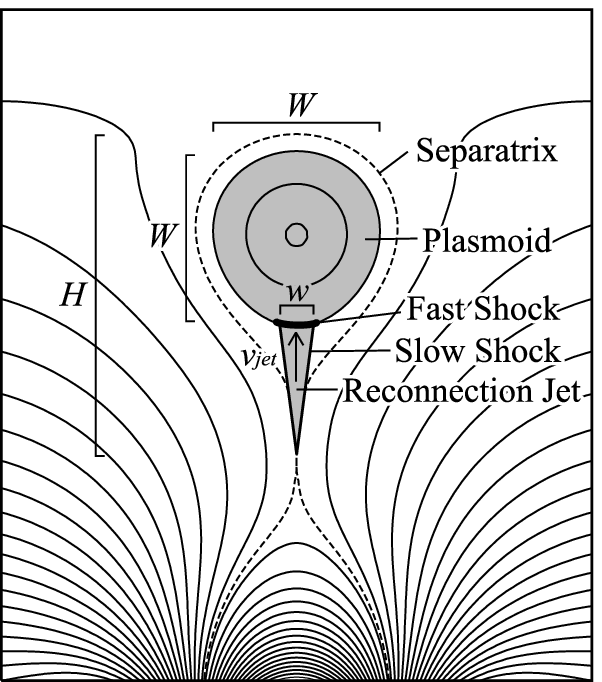}
\caption{Schematic illustration of an acceleration mechanism of a plasmoid. The magnetic field configuration is the result of the simulation at $t=20$. $H (> W)$ is the spatial scale of the magnetic field, which can be consider as the distance between the inflow region and the top of the plasmoid where the magnetic pressure is low.}
\label{figure:acceleration}
\end{figure}

In this appendix, we examine an acceleration mechanism of a plasmoid semi-analytically.
Let us consider the situation shown in Figure \ref{figure:acceleration}.
In this situation, a plasmoid can be accelerated by the momentum added by the reconnection jet and the magnetic pressure gradient force, and decelerated by the magnetic tension force.
Here, the magnetic tension force can be neglected in the magnetic field configuration in Figure \ref{figure:acceleration}, since there is almost no overlying closed field above the plasmoid.
(Note that the gravitational force is neglected since there is no gravity in our simulations, and the gas pressure can also be neglected in the region around the initial plasmoid.)
In this magnetic field configuration,
the force provided by the reconnection jet is estimated as
\begin{equation}
F_{jet}=\rho v_{jet}^2 w \sim \rho v_A^2 w\ \mathrm{[dyn\ cm^{-1}]},
\end{equation}
where $\rho$ is the mass density in the jet, $v_{jet}$ is the velocity of the jet, which is comparable with the local Alfv\'{e}n velocity ($v_A$) in the inflow region,  and $w$ is the width of reconnection jet.
The force exerted over the plasmoid due to is estimated as
\begin{equation}
F_{mag}=W^2 \frac{d}{dy}\frac{B^2}{8\pi} \sim W^2 \frac{1}{H}\frac{B^2}{8\pi}\ \mathrm{[dyn\ cm^{-1}]},
\end{equation}
where $W$ is the spatial size of the plasmoid and $H (> W)$ is the spatial scale of the magnetic field.
The ratio between these forces is as follows:
\begin{equation}
M=\frac{F_{jet}}{F_{mag}}=2\left(\frac{H}{W}\right)\left(\frac{w}{W}\right)
\end{equation}
When $M$ is greater than unity, the momentum provided by the reconnection jet is dominant.
In this work, numerical simulation shows $w/W \sim 1/4$, $H/W \sim 3/2$ to $2$ and $M\sim 1$ so that the these forces are comparable.

%\bibliography{apj-jour,reference}

\end{document}